\documentclass[a4paper,10pt]{article}
\usepackage{url}
\usepackage{helvet}
\usepackage{epsfig}
\usepackage{times}

\usepackage{floatflt}
\usepackage{graphics}
\usepackage{amsmath}
\usepackage{amsthm}
\usepackage{amsfonts}
\usepackage{amssymb}

\begin{document}


\title{Observation, Characterization and Modeling of Memristor Current Spikes
}

\author{Ella Gale~\footnote{e-mail: ella.gale@uwe.ac.uk} $^1$, Ben de Lacy Costello$^1$ and Andrew Adamatzky$^1$\\ 
\\
1. Unconventional Computing Group, University of the West of England,\\
Frenchay Campus, Coldharbour Lane, Bristol, BS16 5SR, UK}

\abstract{Memristors have been compared to neurons (usually specifically the synapses) since 1976 but no experimental evidence has been offered for support for this position. Here we highlight that memristors naturally form fast-response, highly reproducible and repeatable current spikes which can be used in voltage-driven neuromorphic architecture. Ease of fitting current spikes with memristor theories both suggests that the spikes are part of the memristive effect and provides modeling capability for the design of neuromorphic circuits.}


\maketitle

\section{Introduction}

Neuromorphic computing is the concept of using computer components to mimic biological neural architectures, primarily the mammalian brain. 
Although an area of current and active research, we do not know exactly how the brain works, however it is believed that the brain is a neural net. 
Signals travel along neurons via voltage spikes known as action potentials which are caused by the movement of ions across the neuron's cell membrane, and the signals pass between neurons via chemical neurotransmitters (the gap crossed between neurons is the synapse)~\cite{1}. 
The interaction of these spikes is thought to be a cause of brain waves, thought, learning and cognition. The long-term potentiation of neurons is related to a change in structure of the synaptic cleft, which is thought to result from the Spike Time Dependent Plasticity (STDP) of these synapses and result in Hebbian (associative) learning~\cite{2}. 

The memristor is the 4$^{\mathrm{th}}$ fundamental circuit element as predicted by Leon Chua~\cite{3}.  
First reported experimentally using this terminology in 2008~\cite{4}, memristors have been an object of scientific study for at least 200 years~\cite{5}. 
Memristor theory was first demonstrated in a model of the action of nerve axon membranes in 1976~\cite{6}, which was proposed as an alternative to the Hodgkin-Huxley circuit model) and this has led to the suggestion that they would be appropriate components for a computer built using a neuromorphic architecture~\cite{4}. 
Several simulations of neural nets containing memristors have been performed (see for example~\cite{7}). 
Recently, it was reported that circuits combining two memristors with two capacitors could produce self-initiating repeating phenomena similar in form to brain waves~\cite{8}.
 
Perhaps it is not merely the case that memristor models fit neuron behavior, but that neurons themselves are memristive. Thus, we would expect that advances in the study of memristors would explain neurological phenomena (as happened with computer science and STDP). A circuit theoretic analysis of an updated version of Hodgkin-Huxley's model of the neuron has been undertaken~\cite{Chua1,Chua2}. The Hodgkin-Huxley model is often used to explain the transmission of voltage spikes along the neuron. However, this model predicts huge inductances which are not experimentally observed in biology and it has been demonstrated~\cite{Chua1} that updating the Hodgkin-Huxley model with memristors avoids this requirement. A recent paper suggested that memristance could explain the STDP in neural synapses~\cite{2}. The authors used memristor equations to adjust simulated spikes, found a similarity to experimentally measured biological synapse action~\cite{9} and concluded that a memristive mechanism was behind the biological STDP phenomenon. 

In this paper we will show experimentally that memristors spike naturally and do not require a spiking input to cause them to spike in a manner qualitatively similar to neurons. We shall attempt to quantify the spikes. We will then demonstrate that these spikes are also present in theoretical models of memristors and discuss the cause of them. We think that utilizing these naturally-occurring spikes will be the most fruitful way to create neuromorphic memristor architectures.

\section{Properties of Memristor Spikes}

Memristors come in two flavours, charge-controlled (left) and flux-controlled (right) as shown below in Equation~\ref{eqn:1} where $q$ is the charge, $\varphi$, is the magnetic flux, $M$ is the memristance and $W$ is the memductance (inverse memristance)~\cite{3}

$$
V(t) = M(q(t)) I(t) , \: I(t) = W(\varphi(t)) V(t) \, .
\label{eqn:1}
$$

For a charge-controlled memristor we would input a current, $I$, and measure the voltage, $V$. Biological neurons may be described as charge-controlled because it is the movement of ions that causes the change in voltage giving rise to a voltage spike. Our memristors are flux-controlled and a change in voltage causes a spike in the current. Thus, creation of a neuromorphic computer with memristors will be using the complimentary effect to the one utilized by nature, in that memristors have voltage-change-caused current spikes and neurons have current-change-caused voltage spikes. That both types of spikes have a similar form arises from the similarity in the underlying electromagnetics, in that circuits can considered as being constructed with either a voltage source or a current source.

\begin{figure}[htpb]
	\centering
		\includegraphics[width=8cm]{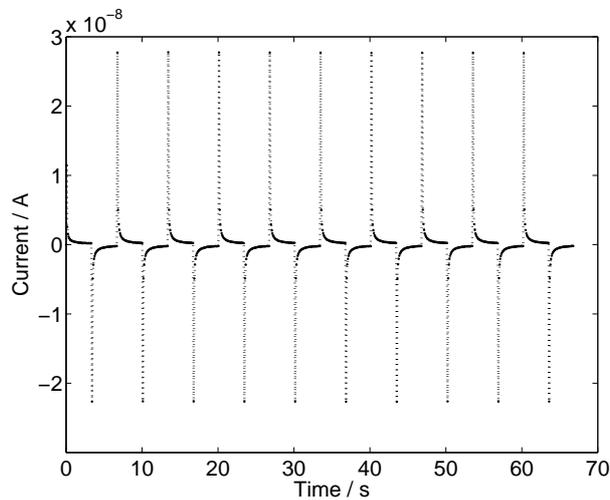}
	\caption{Current spikes recorded from a memristor subjected to the voltage square wave in figure~\ref{fig:Spike7V}. The spike heights are highly repeatable and qualitatively resemble neuronal spikes.}
	\label{fig:Spike7I}
\end{figure}
\begin{figure}[htpb]
	\centering
		\includegraphics[width=8cm]{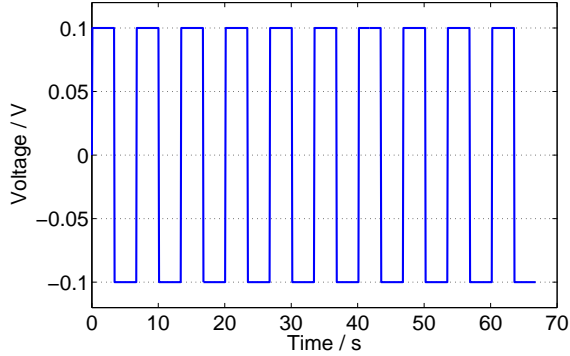}
	\caption{Voltage square wave that the memristor measured in figure~\ref{fig:Spike7I} was subjected to.}
	\label{fig:Spike7V}
\end{figure}

Our memristors are flexible sol-gel titanium dioxide gel layers sandwiched between aluminium electrodes~\cite{10,11} and they show a distinctive large spike that occurs when the voltage is changed. The experiments reported here were carried out with a Keithley 2400 sourcemeter sourcing voltage. There are no spikes in the voltage profiles, (see Figure~\ref{fig:Spike7V}) and no current spikes are seen when the same experiment is done across a resistor. It has been suggested that these spikes are capacitance; however the timescale is too long. The spikes have been reported by other groups in their memristors (see for example, ~\cite{12}), however they are usually overlooked or attributed to artefacts arising from the experimental set-up or not reported at all (many researchers only report the $I-V$ curves to demonstrate that they have a memristor). However, the current spike is an equilibrating process that is responsible for the frequency dependence of the $I-V$ curves. In Figure~\ref{fig:Spike7V} each voltage step had 40 timesteps ($\approx$ 3.3s) to equilibrate. If the voltage is scanned quicker than this, the current has not equilibrated and thus current is higher than the equilbration current. Thus, a faster switching time increases the hysteresis of the $I-V$ loop. This effect increases with frequency until it reaches the limit where the voltage frequency is too fast for the memristor to relax at all and the $I-V$ curve just traces out the maximal spike currents for each voltage.

\begin{figure}[htbp]
	\centering
		\includegraphics[width=8cm]{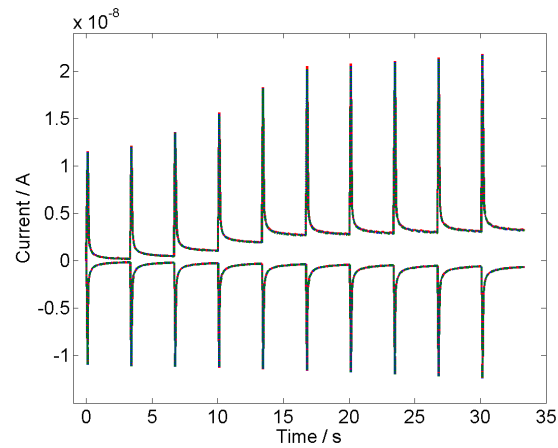}
	\caption{The spikes for 5 successive runs up and then down the voltage staircase shown in figure~\ref{fig:SpikeRepeatsV}. The runs are coloured and overlap. The spikes are highly reproducible on successive runs}
	\label{fig:SpikeRepeats}
\end{figure}

\begin{figure}[htbp]
	\centering
		\includegraphics[width=8cm]{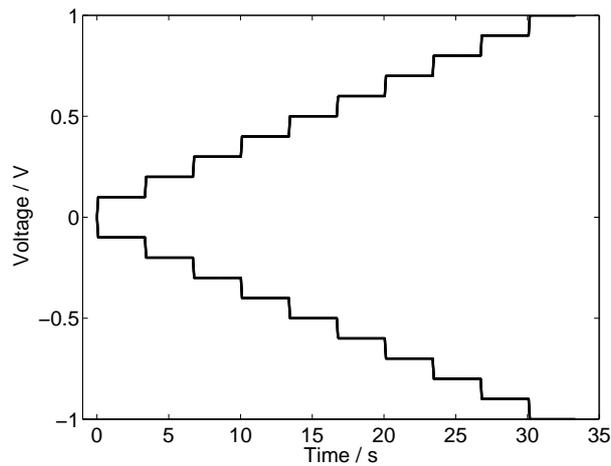}
	\caption{Voltage ramps for figure~\ref{fig:SpikeRepeats}. 5 sets of positive voltage ramps-negative voltage ramps were run, to give the spike response in figure~\ref{fig:SpikeRepeats}.}
	\label{fig:SpikeRepeatsV}
\end{figure}

These current spikes can be seen whenever a voltage change occurs across the memristor. Unlike some neuronal spikes, the voltage does not need to spike. The current spikes are highly reproducible. For the experiment shown in Figure~\ref{fig:Spike7I} (10 pairs of positive to negative switches), the standard deviation was 0.0729\% of the mean for the negative voltages (where $n=10$) and 0.1192\% of the positive voltages (where $n=9$, due to incomplete recording of the first spike)). For the repeated spikes in figure~\ref{fig:SpikeRepeats} (3 repeats each of both positive and negative ramps, as shown in figure~\ref{fig:SpikeRepeatsV}) the largest difference between the spike current repeats was only 3.06$\times10^{-9}$A and only 2.33$\times10^{-10}$A for the equilibrated current - both taken from the positive side as it has a larger hysteresis than the negative side. 

The direction of the current spikes is related to the change in voltage, not its sign, so a change from a positive voltage to zero (turning the voltage source off) gives a negative spike and vice versa for a negative voltage to zero. The spike current still flows for a short while after the voltage source has been turned off. This lag is a general thing and has been recorded in several different devices. In different devices the spikes are the same shape and seem to be following similar dynamics. The spike current is proportional to the equilibrated current. Intriguingly, spike shape closely resembles that of Bi and Poo's experimentally observed STDP function~\cite{9} and thus could be used to perform a similar function.

\section{A Mathematical Description of the Spikes}

\begin{figure}[htbp]
	\centering
		\includegraphics[width=8cm]{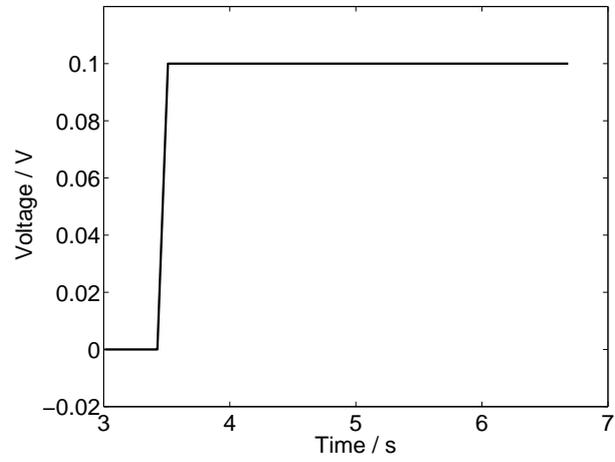}
	\caption{An example of a voltage step as applied to a memristor.}
	\label{fig:VStep}
\end{figure}

\begin{figure}[htbp]
	\centering
		\includegraphics[width=8cm]{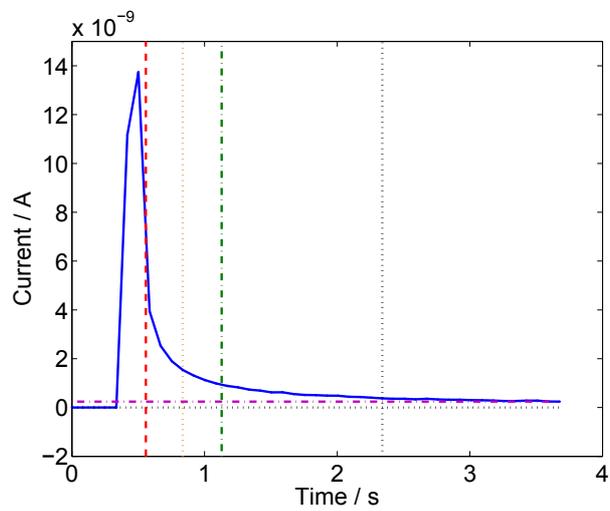}
	\caption{An example Spike. Red dashed line: $\tau_{50}$; orange dotted line $\tau_{90}$; green dot-dashed $\tau_{95}$; blue dotted $\tau_{99}$. Horizonal purple dot-dashed line is $i_\infty$ and the spike height is $i_0$.}
	\label{fig:ExampleSpike}
\end{figure}

\begin{figure}[htpb]
	\centering
		\includegraphics[width=8cm]{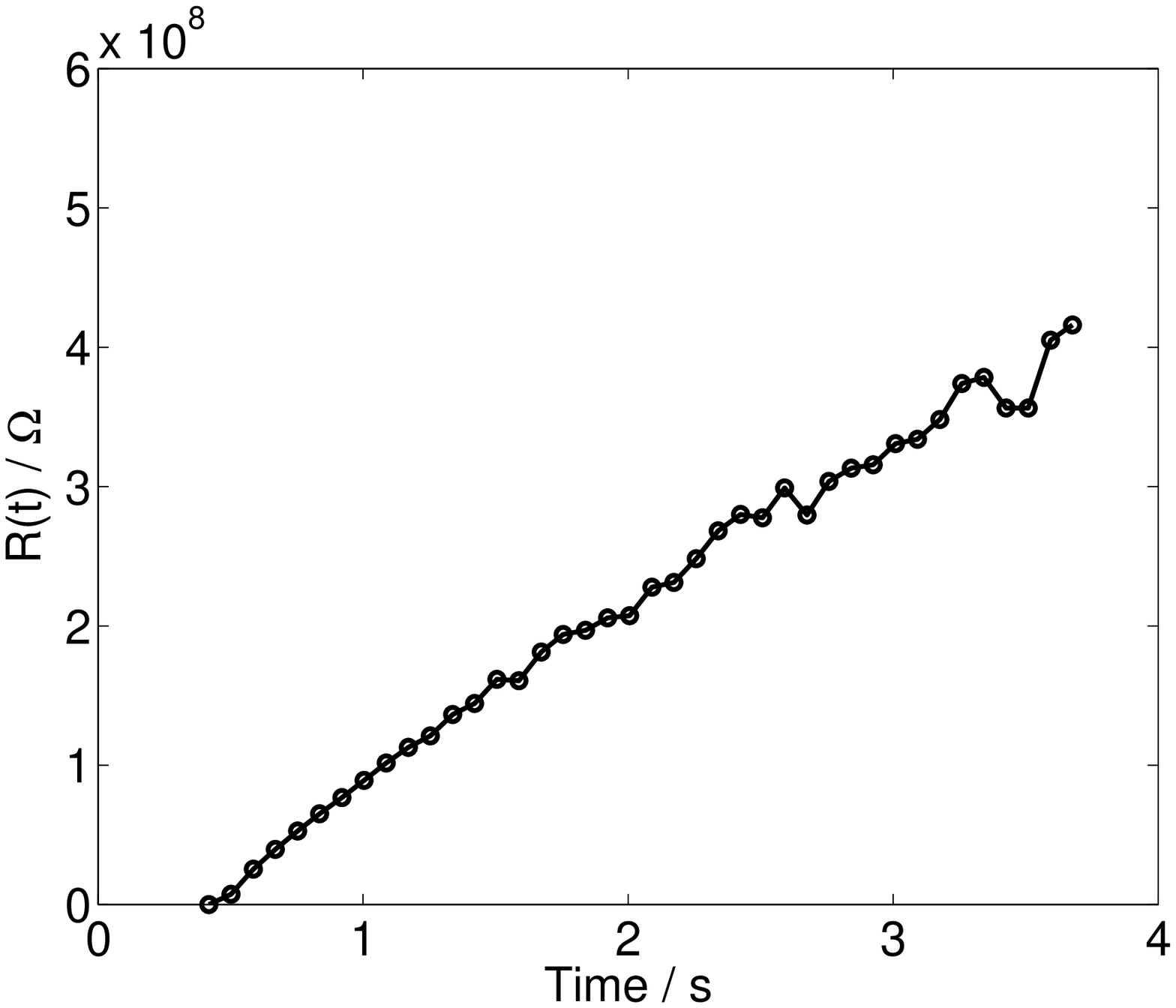}
	\caption{The resistance profile for the memristor subjected to the voltage in figure~\ref{fig:VStep}. Note that the `zero' resistance is due to zero measured resistance as no voltage is applied, not a true zero resistance.}
	\label{fig:ResistanceProfile}
\end{figure}

Figure~\ref{fig:ExampleSpike} shows the $I-t$ response of a single spike to a voltage step like that shown in figure~\ref{fig:VStep}. The current spikes are roughly the same shape, and thus we can make some statements about the nature of the current spikes in memristors, which should also relate to the voltage spike in neurons. As shown in figure~\ref{fig:ExampleSpike}, there is a steady-state current, $i_{\infty}$, a spike current $i_0$ and a transition between the two which is a time-dependent transient $i(t)$. We don't currently know if the $i(t)$ is dependent on $i_0$ or not. We do know that $i_0$ is related to $i_{\infty}$. Until a thorough experimental study is undertaken, we shall assume that $i(t)$ is not dependent on $i_0$ as this is what the experimental evidence seems to suggest. 

Thus, the time-dependent current response, $I(t)$ is assumed to be of the form:
$$
I(t) = i_{\infty} + i(t)
$$
where $i_0  < i(t) < 0$. 

The current response to the voltage is thus:
$$
\Delta I = \frac{V}{R(T)}
$$

The time taken to get to $i(t) = 0$ the equilibration lifetime which we shall call $\tau$, and this lifetime is the short-term memory of the memristor and relates to its dynamical properties; from longer time spike studies with our devices, we know that $\tau$ is approximately 3.3s. We shall define the concept of the equilibration frequency as the `frequency' associated with changing a descretised triangular voltage waveform such that each voltage step $n$ lasts for $\tau$ seconds.

We know that
$$
q_e = \int I(t) dt .
$$
thus, the total measured charge in a memristor spike is 
$$
\Delta q_{\mathrm{spike}} = \int_{t = 0}^{\tau} = i(t) d t + i_{\infty}\tau .
$$
This number includes all the charge carrying species in the system. Knowledge of this number may help us elucidate the mechanism of the spikes. For our example system shown in figures~\ref{fig:ExampleSpike}, we have an $i_{0}$ of 1.37$\times 10^{-8}$A, an $i_\infty$ of $2.40\times 10^{-10}$A, with the $\tau_{50}$ of 0.56s and an $\tau_{90}$ of 0.84s, which shows how quick the fall off is (and $\tau_{95}$ of 1.13s and $\tau_{99}$ of 2.34s, as drawn in figure~\ref{fig:ExampleSpike}). The resistance profile for the memristor subject to a voltage step as shown in figure~\ref{fig:VStep} is shown in figure~\ref{fig:ResistanceProfile}. This is approximately a straight-line which is interesting as it is not required to be by memristor theory and tells us that the spike current response depends on a quantity in the system that is varying with linearly time.

\section{The Mem-Con Theory as Applied to Memristor Spikes}

The mem-con model of memristance~\cite{14} is a recently announced theoretical model that relates real world $q$ and $\varphi$ to Chua's constitutive equations and has been successful in modeling our memristors~\cite{15}. The mem-con theory has the concept of a memory property, the physical or chemical attribute of the device that holds the memory of the device. In titanium dioxide (and many others) it is related to the number of the oxygen vacancies. The presence of oxygen vacancies allows the creation of a doped form of titanium dioxide TiO$_{2-x}$ which is more conducting than the undoped (TiO$_2$) form. The mem-con theory requires that we calculate the memristance from the point of view of the memory property, i.e. the ions.

Theoretically, the voltage step is a discontinuous function and the voltage changes from voltage A, $V_A$ to voltage B, $V_B$ in an infinitesimal, i.e. $\Delta V = \frac{V_B \rightarrow V_A}{t} , t\rightarrow \delta t$. Experimentally this is not the case of course, but the response timescale of the memristor is long enough that we needn't worry about this approximation. 

Thus to elucidate what happens to the memristor during a current spike, and how the final current $i_\infty$ is determined, we take differences of the mem-con theory. We shall assume our device is a TiO$_2$ memristor, with oxygen vacancies acting as the memory property~\cite{14}.

As a reaction to the voltage step, we get a current spike, $\Delta i$, which can be expressed as a volume current within the device as $\Delta J$ as given by:

$$
\Delta \vec{J} = \{ \frac{\Delta q_v \mu_v \vec{L}}{vol}, 0, 0 \}
$$

for vacancies moving in the $+x$ direction where $q_v$ is the charge in that volume due to the vacancies, $\mu_v$ is the ion mobility of vacancies and $L$ is the average electric field causing the movement of the vacancies and $vol$ is the volume full of moving ions. The change in the magnetic field at point $p$, $\Delta \vec{B}(p)$ would then be:

\begin{equation}
\Delta \vec{B}(p) = \frac{\mu_0}{4 \pi} \int \frac{\Delta J \vec{\hat{J}} \vec{\times} \vec{\hat{r}}}{r^{2}} d \tau
\end{equation}

where $\mu_0$ is the permittivity of a vacuum, $\vec{\hat{J}}$ and $\vec{\hat{r}}$ are the unit vectors for $\vec{J}$ and $\vec{r}$ where $\vec{r}$ is the vector of length $r$ from the volume infinitesimal $d \tau$ to point $p$, given by $\vec{r} = \{ r_{x} \hat{\imath}, r_{y} \hat{\jmath}, r_{z} \hat{k} \}$. 





Thus, to get a measure of the effect of the spikes, we need to solve this integral over a time-interval  covering from the start of the spike to the tail-off of the memristor's response. The voltage input is non-integrable, but we can integrate from the start of the step, which we shall take as $t(n)$ where $n$ is the number of the voltage step, which is zero for this case if it is understood that this is not the zero at the start of an experiment with many steps (i.e. we are considering a case as in figure~\ref{fig:ExampleSpike}) to when the memristor has responded, which we shall take as $T$. Dependent on the situation $T$ can be one of many values, for a staircase we would presumably want $T = t(n+1)$ where $t(n+1)$ is the time that the voltage step is input. For a response to a single step function we could take the integral out to $\infty$ (which is what we shall do here). For experimental purposes we might be more interested in integrating to $\tau$ or $\tau_{90}$. 



Solving the integral gives:

$$
\Delta \vec{B}(p) = \frac{\mu_0}{4 \pi} L \mu_{v} \Delta q \{ 0, - x z P_{y}, x y P_{z} \}
$$

with


\begin{eqnarray*}
P_{y} & = & \: \frac{ F}{2 \left( \Delta w^{2} + E^{2} + F^{2} \right)^{\frac{3}{2}}} \\
 & & - \frac{1}{2 \Delta w E F} \frac{\left(
\Delta w E \left( F^{2} \left( E^{2} + F^{2} \right)^{2} +
 a + 
 b \right) \right)}
{c} \\
& & + F \arctan \left( \frac{ \Delta w E }{F \sqrt{\Delta w^2 + E^2 + F^2} } \right) , \\
& \mathrm{and} & \\
P_{z} & = & \: \frac{E}{2 \left( \Delta w^{2} + E^{2} + F^{2} \right)^{\frac{3}{2}}} \\
 & & - \frac{1}{2 \Delta w E F} \frac{\left(
\Delta w F \left( E^{2} \left( E^{2} + F^{2} \right)^{2} +
 a +
 b  \right) \right)}
{c} \\
 & & + E \arctan \left( \frac{ \Delta w F }{E \sqrt{\Delta w^2 + E^2 + F^2} } \right) \, ,
\end{eqnarray*}

where 
$$
a  =  \Delta w^{4} \left( 2 E^{2} + F^{2} \right)
$$
$$
b  =  \Delta w^{2} \left( 2 E^{4} + 5 E^{2} F^{2} + 2 F^{4} \right)
$$
$$
c = \left( \Delta w^2 + F^2 \right) \left(E^2 + F^2 \right) \left( \Delta w^2 + E^2 + \: F^2 \right)^{\frac{3}{2}}.
$$


Where the effect on the magnetic field is due to both the influx of charge and the resulting movement of the boundary between doped and undoped TiO$_2$. 

To calculate the change in magnetic flux through a surface associated with this field, $\varphi$, we need to take the surface integral

$$
\Delta \varphi = \int \Delta \vec{B} \cdotp d \vec{A}
$$

where $d \vec{A}$ is the normal vector from the surface infinitesimal $d A$. 

As it is a surface integral, to calculate the magnetic flux we need to pick a surface to evaluate over. It makes sense to choose a surface that correlates to one of the surfaces of the device. Picking the surface just above the device ($0<x<D$, $0<y<E$, $z=F$), we use the surface normal area infinitesimal, $\vec{d A}$, which is given by $\vec{d A} = \{0,0,\hat{\imath} \hat{\jmath}\}$. As is standard in electromagnetism, we integrate over the entire area. The limits of the surface are taken to be the dimensions of the device.

Thus we derive the general form of the magnetic flux passing through a surface $i$-$j$:
where, because $\varphi$ is entirely dependent on $q$, which is time-varying, we can include the time varying effects by taking the differentials thus
\begin{equation}
	\delta \varphi=\frac{\mu_{0}}{4 \pi} L \mu_{v} i j P_{k} \delta q_{v} \:  ,
	\label{eq:1b}
\end{equation}	

And, as in mem-con theory~\cite{mem-con}, by using Chua's constitutive relation for the memristor, we can then arrive at the change in the Chua memristance as experienced by the ions:

\begin{equation}
	\Delta M_q \left( \Delta q_{v} \left( t \right) \right) =  U X \mu_{v} \Delta P_{k}\left( \Delta q_{v} \left( t \right) \right) \: , 
\label{eq:2}
\end{equation} 
where we have gathered up the constants and explicitly included $P_k$'s dependence on $q_v$.

Equation~\ref{eq:2} can be considered as three separate parts: 

\begin{enumerate}
 \item $U$, the universal constants: $\frac{\mu_0}{4 \pi}$.
 \item $X$, the experimental constants: $D E L$. 
 \item the material variable: $\mu_{v} P_{k}$ (called $\beta$ is ref [!!Mem-Con]), this includes the physical dimensions of the doped part of the device and the drift speed of the dopants. 
\end{enumerate}

Writing out the differences explicitly of equation~\ref{eq:2} we end up with:
$$
M(B) = M(A) + U X \mu_v [ P_{k}(q_B) - P_k (q_A) ] ,
$$

which allows us to calculate how the final Chua memristance from knowledge of the peak and final currents. The Chua memristance is written for the vacancy charge, so to put it into the standard format for the electronic current we need to scale it thus:

$$
R_M = C_M M,
$$
where $R_M$ is the electronic resistance of the doped part of the memristor and $C_M$ is a fitting coefficient.

\subsection{Conservation function}

The memory part of the function only describes the effect of the memristance change on the doped part of the memristor. To cover the other one we use the conservation function, this is most easily expressed in terms of $w(t)$, but $w(t)$ is related to $q(t)$ by

$$
w(t) = \frac{\mu_v L q(t)}{E F v_d}.
$$

Thus, the difference in conservation function, $\Delta R_{\mathrm{con}}$, written as a difference equation is:

$$ 
R_{\mathrm{con}}(B)=  R_{\mathrm{con}}(A) + \frac{\left( D - [w(B) - w(A) ] \right) \rho_{\mathrm{Off}}}{EF} \,
$$

which based on the definition of resistivity and where $\rho_{\mathrm{off}}$ is the resistivity of the undoped part of TiO$_2$.

The mem-con model describes a memristor by being the sum of the memory and conservation functions (both written for the electrons) and this then gives us the following expression for the change in time-varying resistance, $R(t)$, as measured after a change from $V_A \rightarrow V_B$ as:

\begin{eqnarray*}
\Delta R(t)  & = & c_m M(A) + R_{\mathrm{con}}(A) + \frac{\rho_{\mathrm{off}} D}{E F} \\
 & & + c_M U X \mu_v [p_k(q_B(t)) - p_k(q_A(t_T)) ] \\
 & & - \frac{L \rho_{\mathrm{off}} \mu_v [q_B(t) - q_A(t_T) ]}{E^2 F^2 v_d} ,
\label{eqn:Diff}
\end{eqnarray*}

where we have substituted for $w$. This equation has two parts:
\begin{enumerate}
 \item $S$, the time-invarient part, which is:\\
 	 $c_m M(A) + R_{\mathrm{con}}(A) + \frac{\rho_{\mathrm{off}} D}{E F}$
\item $Y$, the time variant part:\\
  $c_M U X \mu_v [p_k(q_B(t)) - p_k(q_A(t_T)) ]$ \\
  $- \frac{L \rho_{\mathrm{off}} \mu_v [q_B(t) - q_A(t_T) ]}{E^2 F^2 v_d} \, ,$
\end{enumerate}
the last two terms which are both dependent on $q$ (remember $p_k$ is dependent on $w$ but can also be written in terms of $q$.

In the above equation~\ref{eqn:Diff} highlights a few subtleties of the model. $p_k$ and $q$ are time-dependent and thus change after the voltage step from $V_A \rightarrow V_B$. If we ask the question of what the difference will be between the equilibrated current at $V_A$ and that at $V_B$, $\Delta R_{A_\infty \rightarrow B_\infty }$ equation~\ref{eqn:Diff} collapses to:

\begin{eqnarray*}
\Delta R  & = & c_m M(A) + R_{\mathrm{con}}(A) + \frac{\rho_{\mathrm{off}} D}{E F} \\
 & & + c_M U X \mu_v [p_k(q_B(\tau)) - p_k(q_A(\tau)) ] \\
 & & - \frac{L \rho_{\mathrm{off}} \mu_v [q_B(\tau) - q_A(\tau) ]}{E^2 F^2 v_d} \, ,
\end{eqnarray*}
which is time invariant and allows us to predict the value of the equilibrated current after a voltage step from the equilibrated current from the step before. 

What if there was previous step in which the device did not equilibrate to $i_\infty$? This would happen if the voltage was changed quicker than $\tau$, i.e. $T$ where $T < \tau$. The $q_A(t_T)$ is not $q_A(\tau)$ and thus needs to be shifted by its value as a proportion of $\tau$. As an example, if we sped the voltage ramps up to 90\% of the equilibration frequency, $q_A$ would be $q_A(\tau_{90})$ and the length of a time step would be $\tau_{90}$. At first glance it might appear that this would merely modulate the starting point for $q_B(t)$, which, at times under $t < \tau$, this would be time dependent. But there is the interaction between $q_B(t)$ and $q_A(t_T)$, the memristor hasn't finished responding to $V_A$ and that response should be mixed in with $V_B$, further complicating predictive efforts.

\section{Modeling Memristor Spikes}

The mem-con model consists of sum of two components: the memory function, $M_e$, and Conservation function, $R_c$. The memory function has a fitting parameter $c_m$ within the model to account for the conversion between the material's resistance as for an oxygen vacancy and as for an electron. The conservation function has the fitting parameter $c_{c}$ which accounts for the resistivity of the undoped material, $\rho_{\mathrm{off}}$, which may not be the same as the bulk titanium dioxide. $R_{\mathrm{on}}$ is the final fitting parameter and relates to the resistivity of the doped material, which is the memristor in the equilibrated state and any resistance in the wires. The fitted equation is 

$$
I(t) = \frac{V}{R_{\mathrm{on}}} - \frac{V}{c_c R_c(t) - c_m M_e(t)} \: .
\label{eqn:fit}
$$

\begin{figure}[htpb]
	\centering
		\includegraphics[width=8cm]{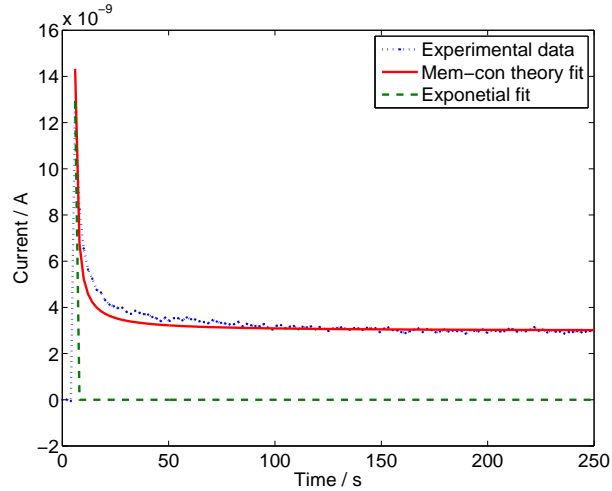}
	\caption{A longer-term spike response fit by the mem-con theory. The mem-con theory fits the experimental data well, the best result fitting the data with an exponential is added as a comparison. Blue dots: experimental data, red line: mem-con fit, green line: exponential fit to the spike.}
	\label{fig:D4aa}
\end{figure}

\begin{figure}[htpb]
	\centering
		\includegraphics[width=8cm]{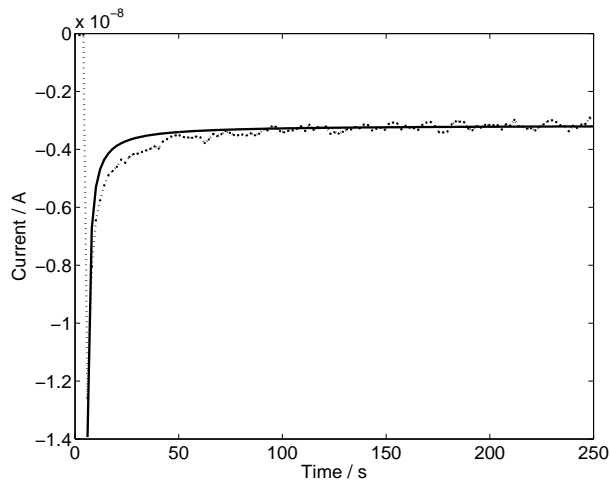}
	\caption{A longer-term negative spike, demonstrating that the negative spikes are fit equally well by the mem-con theory. Blue dots: experimental data, red line: mem-con fit.}
	\label{fig:D4ab}
\end{figure}

As figures~\ref{fig:D4aa} and~\ref{fig:D4ab} shows, the mem-con model fits these spikes quite well and much better than an exponential fit. For the positive spike, $c_M -3.83\times10^{6}$, $c_c = 1.76 \times 10^{6}$ and $V/R_{\mathrm{on}}=2.97 \times 10^{-9}$, with a summed square of residuals of 1.61$\times 10^{-17}$. For the negative spike, $c_M -1.06 \times 10^{6}$, $c_c = 1.86\times10^{-6}$ and $V/R_{\mathrm{on}}=-3.16\times10^{-9}$, with a summed square of residuals of 1.63$\times 10^{-17}$. For the exponential fit, $I(t)=A e^{\lambda t}$, and $A=3.96$, $\lambda=-19.5$ with a summed square of residuals of 2.43$\times 10^{-15}$. The exponential fit could be fit to either the short time spike or the long time tail but not both, the short term spike fit goes erroneously to zero and the long-term spike fit grossly over-estimates the size of the spike. Furthermore, there is no experimental justification for using an exponential fit, unlike the mem-con fit. This model can be utilized to perform simulations of memristor spiking networks to test out possible neuromorphic architectures.

\section{What is the Mechanism?}

The memory property of these memristors is the oxygen ions, usually viewed as positive holes in a semi-conducting material. We suspect that the motion of these ions is behind both the spikes and the memristance as we postulate that the two are the same phenomena. The current that flows at $t=0$s may be the ionic current, which would have a greater inertia, and thus takes longer to stop compared to the electrons, which may explain the cause of the devices hysteresis. This current flow can also explain the open-loop memristors (suggested by Pershin and di Ventra~\cite{16} to explain experimental results such as~\cite{17} which are similar to ones seen in our labs and others'). The spike shape would then be the result of the equilibrating of the ionic current to a change in voltage. We expect that the timescale and dynamics of the spikes will relate to the frequency effects seen in memristors. However, there is much further experimental work to be done to prove this mechanism.

\section{Comparison between memristors and neurons}

Chua's definitions of his two types of memristors, flux  and charge controlled, was given above. The mem-con model has the concept of a two-level system where we have two charge carriers, $q$, our memory property and $e^{-}$ the electronic current which is what is measured in an experiment. Level 0 is the relationship between the vacancy charge, $q$ and vacancy flux, $\varphi$. This is experienced at level 1 by resistance changes ($R(t)$) which effect the electronic current, $I_{e{^-}}$. The circuit measurables are the voltage, $V$ and the total current $I$ where $I = \frac{d q}{d t} + I_{e^{-}}$.

\begin{figure}[htbp]
	\centering
		\includegraphics[width=8cm]{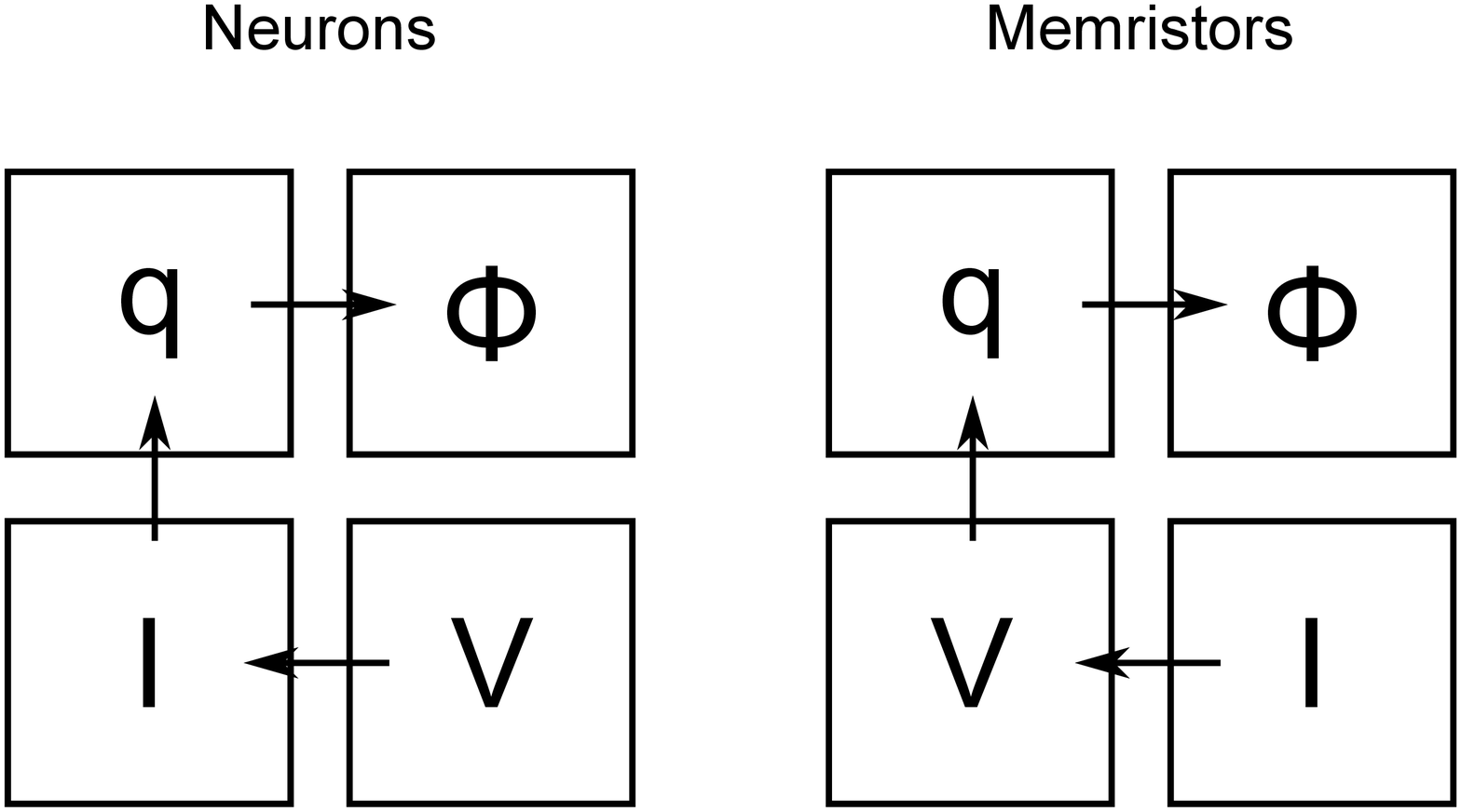}
	\caption{Scheme}
	\label{fig:Scheme}
\end{figure}

For our memristors, driven by a voltage, the right hand side of figure~\ref{fig:Scheme} summarizes the operation. There is a change in voltage, which acts on the electrons and the vacancies, causing a change in the number of charge carriers ($\Delta e^{-}$ and $\Delta q$. The change in $q$ causes a change in the magnetic flux associated with $q$ and thus a change in the Chua memristance. This, due to the conservation of space, causes a change in the amount of material described by the conservation function $R_c$, which then changes the total resistance $\Delta R$. This change in resistance will draw more current, $e^{-}$ and thus the change in the number of electrons is influenced by both the change in voltage and the change in resistance that change has caused. The change in total current is due to both the electrons and the vacancies.

A neuron is the opposite way round, see the left hand side side of figure~\ref{fig:Scheme}. The cell is always pumping ions back and forth, so we have a change current due to an influx of charge carrier. This causes a change in magnetic flux and affects the total resistance (the values of the memory and conservation functions for this system have not yet been worked out). This change in resistance causes a voltage spike. Thus, similarities can been seen between neuronal voltage spikes and memristive current spikes, in that they are the opposite way round with respect to the circuit measurables, in that the memristor as operated here is a current response to a voltage-sourced circuit, and the neuron is a voltage response to a current-sourced circuit. Essentially the shape of the circuit variable, i.e. that which is being measured, is qualitatively similar.

\section{Towards Neuromorphic Computing}

It has been suggested since 1976 that neurons are memristive, but experimental evidence for neuron-like spiking in memristors had not been collated or analyzed in this way before. If this spiking behavior is an integral result of memristance then it is evidence for the suggestion that neurons may be memristive in action and further understanding of memristor theory may further the neurological understanding. 

This work shows that to make neuromorphic computers that compute with spikes memristors are an obvious choice for this task as they spike naturally. Interruption to the equilibrating current curves as shown in figure [!3], by, for example, changing voltage, would potentiate the connection by modifying the memristance and could thus be used to do STDP with memristors without requiring CMOS neurons to generate the spikes.

\section{Conclusion}

Memristors, when subject to a change in voltage, undergo a current spike. This spike has been shown to be reproducible and repeatable. The mem-con theory have been shown to fit the time-dependent current behaviour with only two fitting parameters (which come from the missing material values in the theory) suggesting that this $I-t$ spike behaviour is an aspect of memristance. Rewriting the mem-con theory as a difference equation allows the formulation of a predictive equation to related the equilibrated currents at different (and successive) voltages. Application of the equilibration lifetime ($\tau$) to this equation highlights where the time-responsive interactions might arise in a memristor switched faster than the equilibration frequency.

\section*{Acknowledgement}

This project has been funded by EPSRC grant No. EP/H014381/1.

\begin{center}
\rule{6 cm}{0.02 cm}
\end{center}

\newpage

\
 
\end{document}